\documentclass[letterpaper, 10 pt, conference]{ieeeconf}  

\IEEEoverridecommandlockouts                              

\usepackage{graphics} 
\usepackage{epsfig} 
\usepackage{url} 
\usepackage[noadjust]{cite}

\title{\LARGE \bf
A Scalable and Adaptable Multiple-Place Foraging Algorithm for Ant-Inspired Robot Swarms*
}

\author{Qi Lu$^{1\S}$, Melanie E. Moses$^{1, 2}$, and Joshua P. Hecker$^{1}$ 
\thanks{*This work was supported by a James S. McDonnell Foundation Complex Systems Scholar Award, and NASA MUREP $\#$NNX15AM14A funding for the Swarmathon.}
\thanks{$^{1}$The Department of Computer Science,
        University of New Mexico, Albuquerque, NM 87106, USA.}%
\thanks{$^{2}$External Faculty, Santa Fe Institute, Santa Fe, NM 87501, USA.}%
\thanks{$^{\S}$Correspondence: lukey11@cs.unm.edu}%
}

\begin{document}

\maketitle
\thispagestyle{empty}
\pagestyle{empty}

\begin{abstract}
Individual robots are not effective at exploring large unmapped areas. An alternate approach is to use a swarm of simple robots that work together, rather than a single highly capable robot. The central-place foraging algorithm (CPFA) is effective for coordinating robot swarm search and collection tasks. Robots start at a centrally placed location (nest), explore potential targets in the area without global localization or central control, and return the targets to the nest. The scalability of the CPFA is limited because large numbers of robots produce more inter-robot collisions and large search areas result in substantial travel costs. We address these problems with the multiple-place foraging algorithm (MPFA), which uses multiple nests distributed throughout the search area. Robots start from a randomly assigned home nest but return to the closest nest with found targets. We simulate the foraging behavior of robot swarms in the robot simulator ARGoS and employ a genetic algorithm to discover different optimized foraging strategies as swarm sizes and the number of targets are scaled up. In our experiments, the MPFA always produces higher foraging rates, fewer collisions, and lower travel and search time compared to the CPFA for the partially clustered targets distribution. The main contribution of this paper is that we systematically quantify the advantages of the MPFA (reduced travel time and collisions), the potential disadvantages (less communication among robots), and the ability of a genetic algorithm to tune MPFA parameters to mitigate search inefficiency due to less communication.


\end{abstract}

\section{INTRODUCTION}
\label{section:intro}

A large number of simple individual robots working together has the potential to be useful for tasks which a traditional single expensive, specialized and complicated robot is not able to handle, such as searching in large unmapped areas~\cite{Stormont05}, distributed contaminant cleanup, and rescue~\cite{Kantor2006}. Robot swarms can also be involved in sophisticated problem solving, including cooperative transportation, de-mining, and space exploration~\cite{Brooks89, Landis2004, Fink2005, Karl2016}.

We focus on developing a scalable, decentralized search-and-collection algorithm based on ant-like foraging~\cite{HarvesterAnt1996, Winfield2009, Liu:2010}. The swarm can adapt to changes in swarm size and the number of targets through real-time response to conditions without external or off-line intervention. Each robot in the swarm makes real-time in-situ decisions on whether to communicate using pheromones, forego communication but individually return to search a location, or return to the collection zone to gather additional information from other robots. The robot behaviors are modeled after those of a particular genus of desert seed harvester ants that~\cite{Flanagan2011, Flanagan2012} are restricted to foraging in short-time windows during which not all available targets can be collected; So they are designed to collect as many targets as possible, but not for optimal complete collection~\cite{Joshua2015}. 

Here, we present the multiple-place foraging algorithm (MPFA) with multiple nests that robots depart from and return to. The robots make on-line decisions to switch to new collection zones based on proximity to their last-found target. The MPFA was presented in our recent work~\cite{Qi2016} and it shown that distributing 2, 4, or 8 nests in the MPFA produce higher foraging rates and lower average travel time compared to the central-place foraging algorithm (CPFA) developed by Hecker and Moses~\cite{Hecker:2012, Hecker:2013}. Here we compare the scalability and adaptation of the MPFA to the CPFA when increases the number of robots and the number of targets. In the MPFA we deploy 4 nests uniformly in the same size of search arena. A set of real-valued parameters specifying the individual robot controllers is evolved by a genetic algorithm (GA) to optimize the foraging strategy in the multi-physics robot simulator Autonomous Robots Go Swarming (ARGoS)~\cite{ARGoS}. Every robot in the swarm uses the same controller.  

We evolve foraging strategies for different swarm sizes (4, 8, 16, 32 and 64) and number of targets (128, 256, 512, 1024 and 2048). We observe the average foraging rate, collision time, travel and search time change as swarm size and the number of targets increase.

The remainder of this paper is organized as follows. Section~\ref{section:related} introduces related work. The design of the MPFA and the description of evolution are provided in Section~\ref{section:design} and Section~\ref{section:GA}. The configuration of the MPFA in ARGoS and the experimental results are in Section~\ref{section:experiments} and Section~\ref{section:results}. Section~\ref{section:discussion} discusses the conclusions.

\section{RELATED WORK}
\label{section:related}

Central-place foraging is commonly studied in swarm robotics~\cite{swarmrobotics2008, Brambilla2013}. Hecker and Moses utilized and formalized the behaviors from Flanagan and Letendre's ant field studies~\cite{Flanagan2011, Flanagan2012, Letendre:2013} to create the CPFA. The algorithm is well designed and applied to real physical robots, which are designed on the iAnt robots platform~\cite{Moses2014}. The error-tolerance, flexibility, and scalability were evaluated on both simulated and real robot swarms~\cite{BeyondPherom2015}. However, the simulated robots were not physics-based and collisions between robots were not considered.

The studies on task allocation by Hsieh et al~\cite{Hsieh2008, Halasz2007Dynamic, Berman2008} showed that a bio-inspired approach to the deployment of a homogeneous swarm of robots to multiple sites. The robots autonomously redistribute themselves among the candidate sites to ensure task completion by optimized stochastic control policies. It models the swarm as a hybrid system where agents switch between maximum transfer rates and constant transition rates. In our method, the robots are distributed and initialized to multiple nests evenly. Then, they transit to other nests autonomously based on the distribution of remainder targets and the evolved search strategy. The search strategy is evolved by GA automatically and all the robots use the same strategy.    

There are few studies on multiple-place foraging in biological systems. The polydomous colonies of Argentine ants are comprised of multiple nests spanning hundreds of square meters~\cite{Fast2013, Schmolke09}. A study by Chapman et al~\cite{MCPF1989} showed that a community of spider monkeys can be considered as multiple central place foragers (MCPF). They select a sleeping site close to current feeding areas, and the MCPF strategy entails the lowest travel costs. A study by Tindo et al~\cite{EEN:EEN966} showed that wasps living in multiple nests have greater survival rate and increased productivity. However, multiple-place foraging has not been systematically compared to central placed foraging in robotic swarms which we do here.


\section{THE DESIGN OF THE MPFA}
\label{section:design}
In the MPFA, robots are evenly distributed around nests. They start from a nest but return to the closest nest to their position after finding a target or giving up the search. The use of multiple collection points is the fundamental difference between the CPFA and the MPFA; all other components of the two foraging algorithms are kept deliberately identical in order to test for the effect of multiple nests on swarm foraging efficiency. 

\begin{figure}[thpb]
      \centering
        \framebox{\parbox{3in}{\includegraphics[scale=.40]{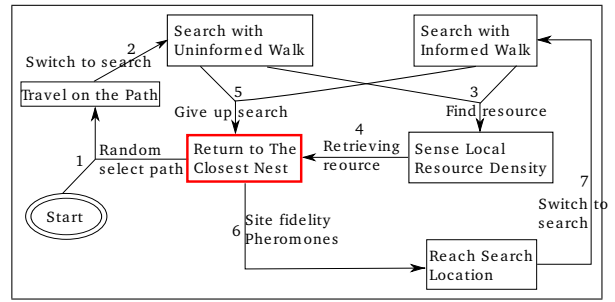}}}
      \caption{The flow chart of an individual robot's behavior in MPFA during an experiment.}
      \label{fig:MPFA_diagram}
\end{figure}

The behavior of an individual robot in an MPFA foraging round is shown in Fig.~\ref{fig:MPFA_diagram}. Each robot transitions through a series of states as it forages for targets. This differs from the CPFA~\cite{BeyondPherom2015} in how the robots return to nests which are in steps 4 and 5. In the MPFA, robots initially disperse from the nests closest to them, followed by random selected travel paths (step 1). An uninformed correlated random walk is used to search targets when robots stop to follow the paths (step 2)~\cite{Fewell1990}. Robots navigate home to nests closest to them when they retrieve targets or give up search (step 4 and 5)~\cite{Crist1991}. Robots that find targets will detect the local target density before return to nests (step 3)~\cite{Holldobler1976}. Robots that are more likely to return to previously found sites using site fidelity or pheromone recruitment (step 6), then they search the sites thoroughly with informed walk (step 7).       

\begin{figure}[thpb]
      \centering
      \framebox{\parbox{3in}{\includegraphics[scale=.27]{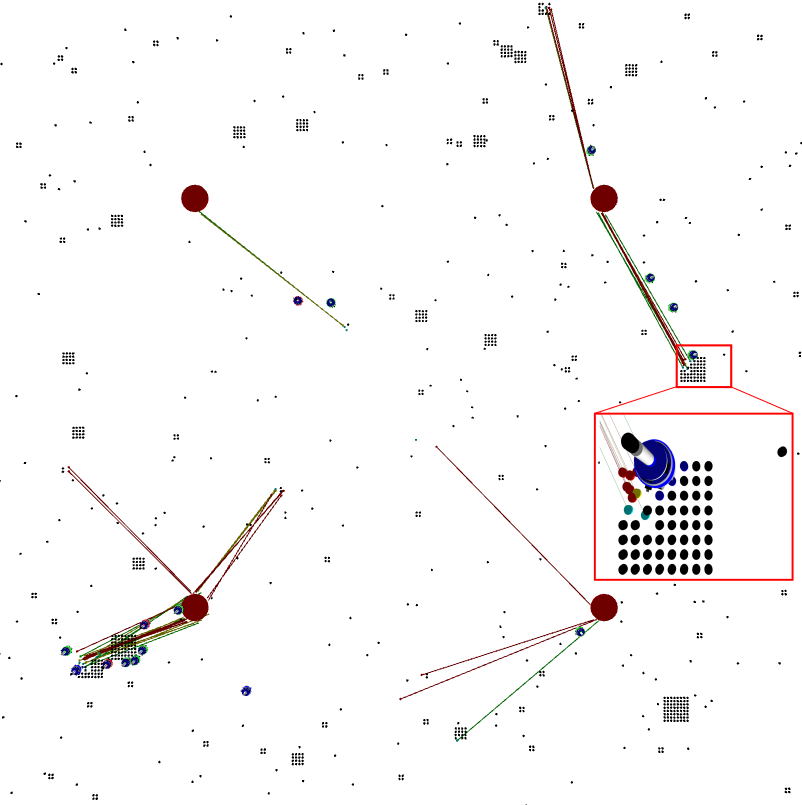}}}
      \caption{The placement of nests and targets in ARGoS. 1024 targets (black points) and 16 robots (larger blue points) are placed in a $15\times15m^2$ arena, 4 nests (red circles) are distributed uniformly in the arena. The targets are arranged in a partially clustered distribution. Colored lines indicate pheromone trails with different strength. A small area is magnified to show a robot, colored pheromone waypoints, a large cluster of targets, and a single target.}
      \label{fig:MPFA_ARGoS}
\end{figure}

In our design, the robots search globally just as in the CPFA -- they can travel in the entire arena. As in the CPFA, pheromone trails are simulated using a list of pheromone-like waypoints to identify target-rich areas. When a robot returns to a nest, it will probabilistically select a waypoint from the nest's list and travel to that location. It shares information (pheromone waypoints) locally at its current nest (see Fig.~\ref{fig:MPFA_ARGoS}). In contrast to the CPFA, pheromone waypoints are globally available to all robots.  

Since robots always return to the closest nest with a found target, the sensed information relevant to a given target neighborhood is always associated with the nest closest to the position of the identified neighborhood. Thus, if a robot follows a pheromone waypoint from a nest, then the distance from the nest to the destination of the pheromone must be the shortest distance to the target neighborhood identified by the pheromone.

\section{The GA EVOLUTION}
 \label{section:GA}
We implement the CPFA and MPFA on a foraging task for different experiments in ARGoS. Furthermore, we use a GA to identify MPFA parameters that maximize foraging strategy. We implement our GA using GAlib~\cite{GAlib} following parameters described by Hecker and Moses~\cite{BeyondPherom2015}. The set of seven MPFA parameters is identical to the set of CPFA parameters. The movement, sensing, and communication of each single robot are evolved and evaluated. The parameters are described in the following,
\begin{itemize}
\item \textbf{Probability of switching to search:} The robot has the probability of switching from travel to uninformed random search. The probability is initialized from a uniform random distribution, $\mathcal{U}(0, 1)$.  
\item \textbf{Probability of returning to nest:} The robot has the probability of giving up search and returning to nest. It is initialized from a uniform random distribution, $\mathcal{U}(0, 1)$.
\item \textbf{Uninformed search variation:} If the robot searches using a correlated uninformed random walk, the successive turning angles $\theta_t$ is defined by $\theta_t=\mathcal{N}(\theta_{t-1}, \sigma)$, where $\theta_{t-1}$ is the turning angle in the current step, and $\sigma$ is the standard deviation or uninformed search variation, which determines the turning angle of the next step. $\sigma$ is initialized from a uniform random distribution, $\mathcal{U}(0, \pi)$. 
\item \textbf{Rate of informed search decay:} If the robot searches using an informed correlated random walk, the standard deviation of the successive turning angles $\sigma$ decays as a function of time $t$, $\sigma = \omega+(2\pi-\omega)e^{-\lambda_{id}t}$, where $\lambda_{id}$ is the rate of informed search decay. $\lambda_{id}$ is initialized from an exponential decay function $exp(5)$.
\item \textbf{Rate of laying pheromone and rate of site fidelity:} The information decisions are governed by parameterization of a Poisson cumulative distribution function as defined by $\mathcal{POIS}(k,\lambda)$, where $k$ is the likelihood of detecting at least $k$ additional resources, and $\lambda$ is the rate of laying pheromone or the rate of site fidelity. It is initialized from a uniformed random distribution, $\mathcal{U}(0, 20)$. The robot returns to a previous location if $\mathcal{POIS}(k, \lambda)>\mathcal{U}(0,1)$. If $k$ is large, the robot is likely to return to the same location using information on its next foraging trip.
\item \textbf{Rate of pheromone decay:} The pheromone decays exponentially over time $t$ as defined by $e^{-\lambda_{pd}t}$, where $\lambda_{pd}$ is the rate of pheromone decay. It is initialized from an exponential decay function $exp(10)$.

\end{itemize}
We repeat the evolutionary process 10 times for the CPFA as well as for the MPFA, in order to generate 10 independently evolved foraging strategies for each experimental configuration. 

In summary, using a swarm size of 40 robots, we evaluate each swarm 8 times on different random placements of targets in the partially clustered distribution to determine their fitness. We use a $50\%$ uniform crossover rate and a $5\%$ Gaussian mutation rate with a standard deviation of $0.02$. We use elitism to keep the individual with the highest fitness.

We altered the termination criteria of the GA in order to hasten parameter convergence and ran the GA for a maximum of 100 generations. The GA terminates based on three criteria: the number of generations, the convergence of fitness, and the diversity of swarm sizes, which are introduced in GAlib~\cite{GAlib}. The GA will stop if the fitness is convergent and the diversity of the population is low. Otherwise, it will stop after 100 generations. Our code is available on GitHub\footnote{\url{https://github.com/BCLab-UNM/iAnt-ARGoS/tree/lukey_development}}. 

In our GA, $89\%$ of the evolution terminates on the convergence of fitness and the diversity of swarm sizes. Across 10 independent evolutionary runs, all evolved parameter sets were nearly equally fit: The standard deviation in fitness was at most $5\%$ of the mean fitness value. The fitness of the best parameter set, evaluated on 100 target placements, is shown in Fig.~\ref{fig:scalability_forage} and Fig.~\ref{fig:adaptation_forage}.


\section{EXPERIMENTAL CONFIGURATION IN ARGoS}
\label{section:experiments}

Table~\ref{table:configuration} shows the experimental configuration in ARGoS. To test scalability, the number of targets is always 1024, and the number of robots is scaled to be 4, 8, 16, 32 or 64.  We set different foraging time windows for each swarm, depending on the swarm size. The selected times allow the evolved swarms to collect approximately half of the targets. The foraging time of robots are the same across all experiments: by multiplying the number of robots by the foraging time, we have 480 robot-minutes (or 8 robot-hours) in our experiments (see Table~\ref{table:configuration}).  

To test adaptation, the number of robots is always 32. The number of targets is 128, 256, 512, 1024 or 2048. The foraging time is set independently for each experiment so that approximately $40\%$ of the targets are collected by the best evolved strategy. All experiments are replicated 100 times. The locations of targets and robots are initialized randomly in the 100 replicates.

\begin{table}[thpb]
\caption{Experimental configuration in ARGoS}
\label{table:configuration}
\begin{center}
\begin{tabular}{|c|c|c|}
\hline
 & Robots & $4$, $8$, $16$, $32$ or $64$\\
\cline{2-3}
Scalability & Targets & $1024$ \\
\cline{2-3}
& Time (minutes) & $120$, $60$, $30$, $15$ and $7.5$\\
\hline
 & Robots & $32$\\
\cline{2-3}
Adaptation & Targets & $128$, $256$, $512$, $1024$ or $2048$ \\
\cline{2-3}
& Time (minutes) & $5$, $8$, $10$, $12$ and $30$\\
\hline
\end{tabular}
\end{center}
\end{table}

The targets are placed in a partially clustered distribution. This distribution has various sizes of square clusters. The targets are placed either in a large cluster, a medium cluster or individual targets in a uniform random distribution (see Fig.~\ref{fig:MPFA_ARGoS}). Both algorithms are tested in a simulated arena size of $15\times15m^2$. The CPFA has one center nest and the MPFA has 4 uniformly and evenly distributed nests. 

\section{RESULTS}
\label{section:results}
We compare the efficiency of the CPFA and the 4 nest MPFA on foraging rate, collision time, and travel and search time when the swarm sizes and the number of targets are scaled up. We identify statistical differences using a t-test, and we identify whether performance varies systematically by calculating a log-linear regression in which the performance are compared to the $\log_2$ of the swarm sizes or the number of targets. 

\subsection{Foraging Efficiency}

\begin{figure}[thpb]
      \centering
        \framebox{\parbox{3in}{\includegraphics[scale=.40]{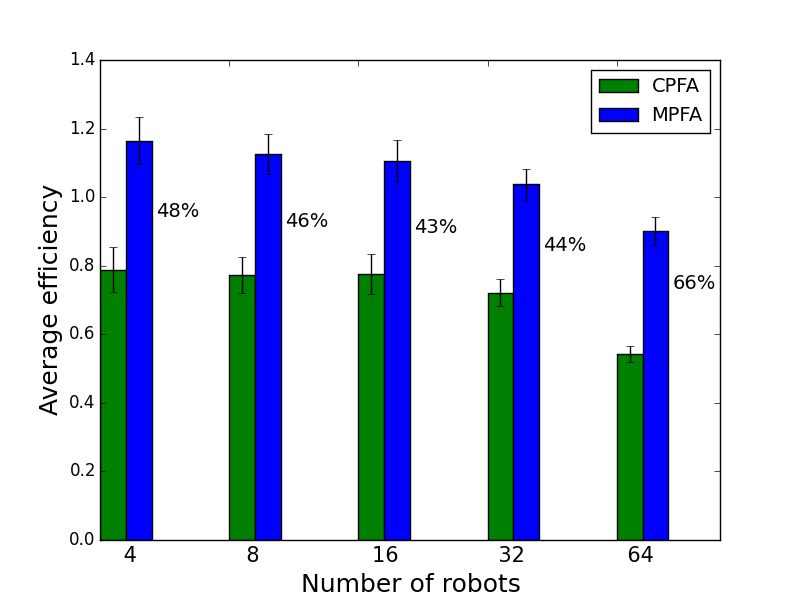}}}
      \caption{The average efficiency (targets collected per robot, per minute) for the CPFA ($p=0.08$) and MPFA ($p=0.04$) decrease as the swarm size increases. The $p$ value is from the average of collected targets and the $\log_2$ of the swarm size. Results are for 100 replicates. The percentage of improvement is labelled.}
      \label{fig:scalability_forage}
\end{figure}

The total foraging rate of each swarm is the sum of the total collected targets in the swarm. We measure the average foraging rate, which is the number of targets per robot collected in every minute. Fig.~\ref{fig:scalability_forage} shows the average foraging rate as the swarm size increases. The average foraging efficiency of the MPFA exceeds that of the CPFA in all cases, by up to $66\%$ in the case of 64 robots.

\begin{figure}[thpb]
      \centering
        \framebox{\parbox{3in}{\includegraphics[scale=.40]{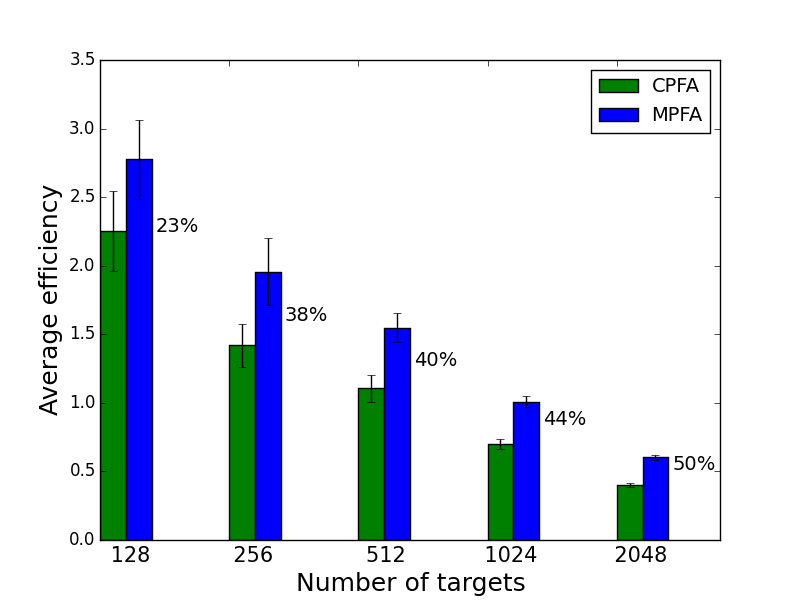}}}
      \caption{The average efficiency (targets collected per robot, per minute) for the CPFA ($p=0.04$) and MPFA ($p=0.001$) decrease as the number of targets increases. The $p$ value is from the average of collected targets and the $\log_2$ of the number of targets. The efficiency is always higher for the MPFA.}
      \label{fig:adaptation_forage}
\end{figure}

Fig.~\ref{fig:adaptation_forage} shows the average foraging rate as the number of targets increases. The average foraging efficiency of the MPFA exceeds that of the CPFA in all cases, by up to $50\%$ in the case of 2048 targets.
 

\subsection{Collision Efficiency}

In our simulation, if the distance between two robots is less than 0.25m, each robot will detect a collision. Each robot senses the location of the other and turns left or right in order to avoid a collision, moving approximately 8 cm before resuming traveling.

\begin{figure}[thpb]
      \centering
        \framebox{\parbox{3in}{\includegraphics[scale=.40]{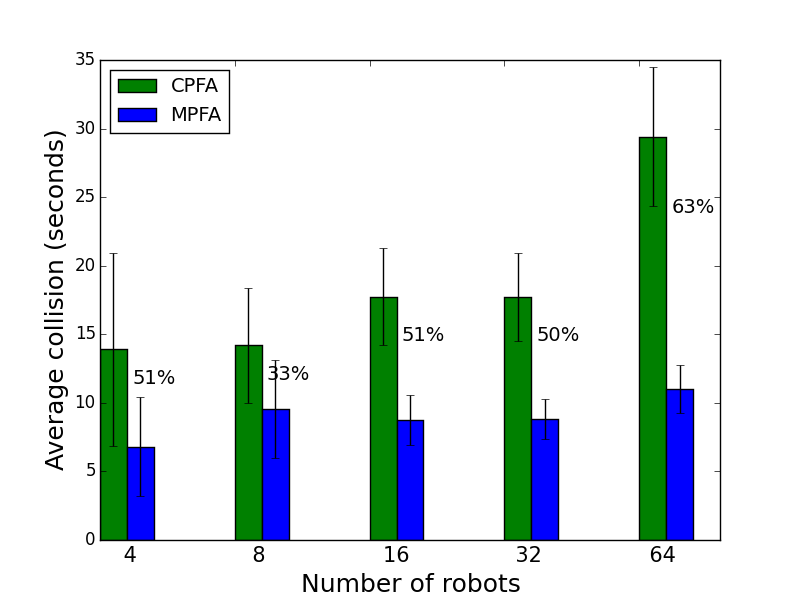}}}
      \caption{The average efficiency (collision time per robot, per target) for the CPFA ($p=0.06$) and MPFA ($p=0.10$) increase as the swarm size increases.}
      \label{fig:scalability_collision}
\end{figure}

The collision time is the time required to avoid a collision. The total collision time of each swarm is the sum of the total collision time for all robots in the swarm. We measure the average collision time, which is the collision time per robot in collecting a target. The "per robot, per target" collision makes the comparison fairly. For "per robot", it is obvious that a larger swarm results in more collisions, but the rate of increase is not obvious. It is easier to analyze the trend of collision rates on each robot rather than on the swarm when the swarm sizes are different. For "per target", calculating 'per target' definitely dilute the comparison. However, the collision is higher if more targets are collected. We should consider the collision rate based on the foraging rate (In our results, the MPFA always has higher foraging rate). The average collision time as swarm size increases is shown in Fig.~\ref{fig:scalability_collision}. 

\begin{figure}[thpb]
      \centering
        \framebox{\parbox{3in}{\includegraphics[scale=.40]{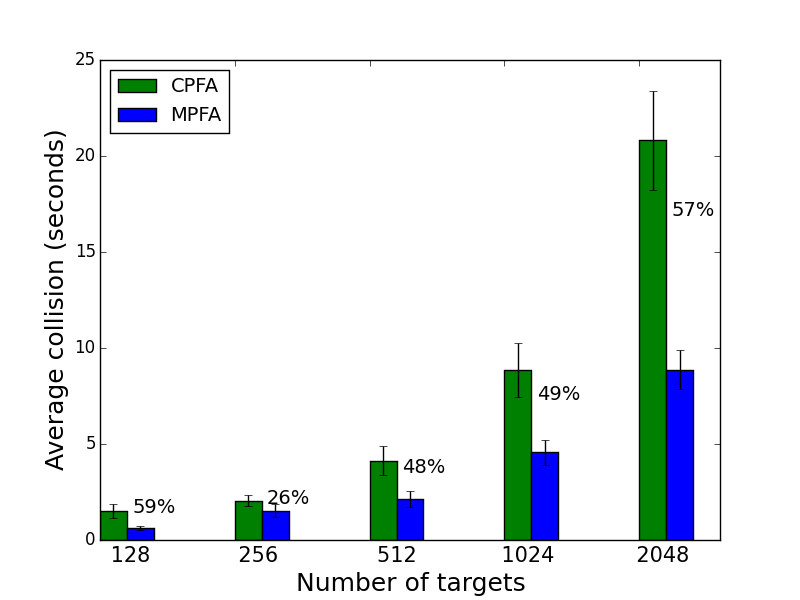}}}
      \caption{The average efficiency (collision time per robot, per target) for the CPFA and MPFA as the number of targets increases ($p=0.03$).}
      \label{fig:adaptation_collision}
\end{figure}

The collision time for the MPFA is less than the collision time for the CPFA. We also see that the collision time for the MPFA is reduced as the number of targets increases (see Fig.~\ref{fig:adaptation_collision}). 

\subsection{Travel and Search Efficiency}

Foraging time is composed of two distinct activities. When a robot departs from its nest, it travels to a location where it starts to search for targets. Once at the destination, the robot engages in a localized search. Once a target is discovered, the robot takes approximately the same \emph{travel time} back to the nest. Some robots take approximately the same travel time back to the location of the discovered target if they are recruited by pheromones, but their \emph{search time} is reduced.

\begin{figure}[thpb]
      \centering
        \framebox{\parbox{3in}{\includegraphics[scale=.40]{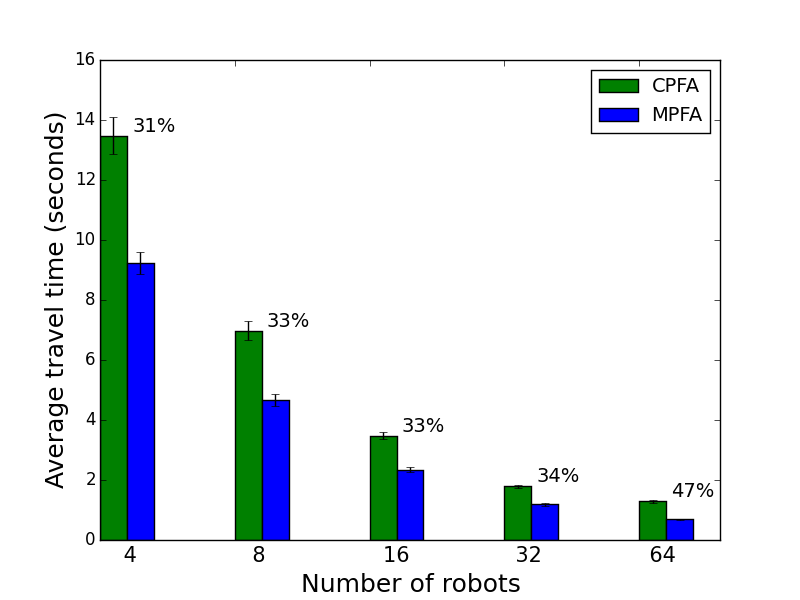}}}
      \caption{The average travel time (per robot, per target) for the CPFA and MPFA decrease as the swarm size increases ($p=0.04$).}
      \label{fig:scalability_travel_time}
\end{figure}

We measure the average travel time and search time spent to collect one target by a robot. The average travel time for the MPFA (see Fig.~\ref{fig:scalability_travel_time}) is less than the CPFA for all swarm sizes. 

\begin{figure}[thpb]
      \centering
        \framebox{\parbox{3in}{\includegraphics[scale=.40]{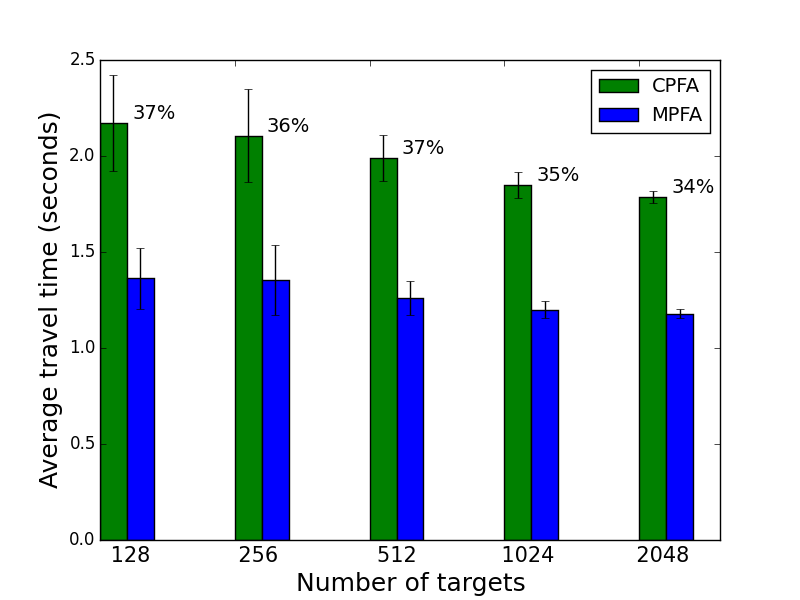}}}
      \caption{The average travel time (per robot, per target) for the CPFA ($p=0.001$) and MPFA ($p=0.03$) as the number of targets increases.}
      \label{fig:adaptation_travel_time}
\end{figure}

The travel time for the MPFA (see Fig.~\ref{fig:adaptation_travel_time}) is also less than the CPFA as the number of targets increases. 

\begin{figure}[thpb]
      \centering
        \framebox{\parbox{3in}{\includegraphics[scale=.40]{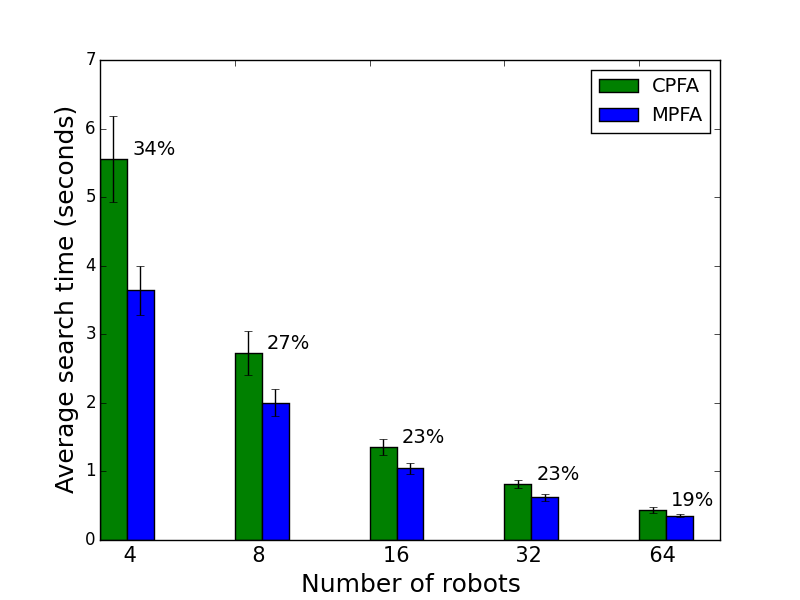}}}
      \caption{The average search time (per robot, per target) for the CPFA and MPFA as the swarm size increases ($p=0.03$).}
      \label{fig:scalability_search_time}
\end{figure}

Fig.~\ref{fig:scalability_search_time} shows that the average search time decreases as the number of robots increases. The search time for the MPFA is less than the CPFA. The search time for the CPFA decreases faster than the MPFA. The improvement is up to $34\%$ in the first case for 4 robots and it is down to $19\%$ in the last case for 64 robots.

\begin{figure}[thpb]
      \centering
        \framebox{\parbox{3in}{\includegraphics[scale=.40]{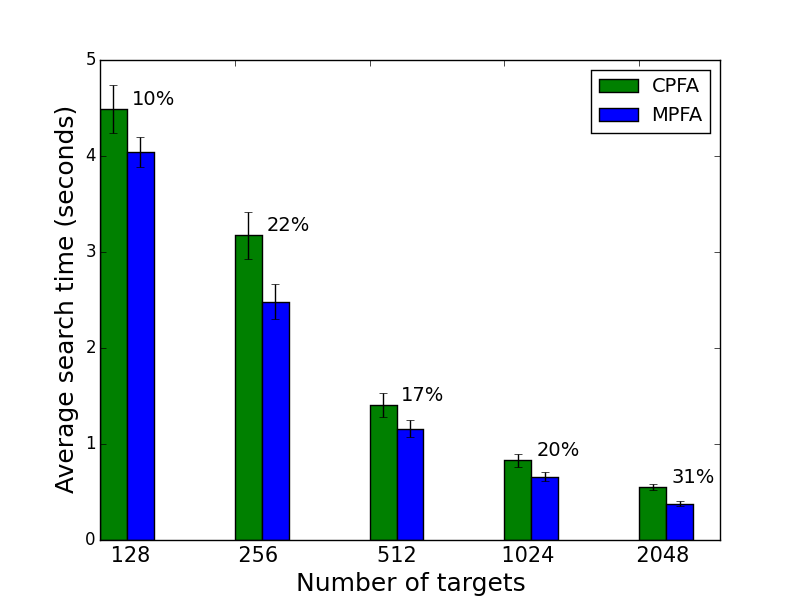}}}
      \caption{The average search time (per robot, per target) for the CPFA and MPFA as the number of targets increases ($p=0.05$).}
      \label{fig:adaptation_search_time}
\end{figure}

The search time decreases as the number of targets increases (see Fig.~\ref{fig:adaptation_search_time}). The search time for the MPFA decreases faster than the CPFA. The improvement goes up to $31\%$ in the last case.

\section{DISCUSSION}
\label{section:discussion}
This paper explores how swarm size and the number of targets affect foraging rates, collision time, travel and search time. Not surprisingly, increasing the swarm sizes or the number of targets decreases the average foraging rate (see Fig.~\ref{fig:scalability_forage} and Fig.~\ref{fig:adaptation_forage}), but decreases slower for the MPFA. This implies that the MPFA is more efficient in larger swarms or in an environment with more targets. 

The average collision time for the MPFA is much less than the CPFA as the swarm size or the number of targets increases (see Fig.~\ref{fig:scalability_collision} and Fig.~\ref{fig:adaptation_collision}). The collision time for the CPFA increases faster as the number of targets increases (see Fig.~\ref{fig:adaptation_collision}). We hypothesize that the more targets there are, the harder robots will spread out in the CPFA. It demonstrates that the MPFA has the advantage of avoiding collisions in large swarm size or in an environment with a large number of targets.  

The increase of swarm sizes makes the average travel time for the MPFA decrease faster than the CPFA (see Fig.~\ref{fig:scalability_travel_time}). This shows that the MPFA has the advantage of reducing travel time as the swarm size increases. We hypothesize that the evolved probability of returning to nest increases faster as the swarm size increases. The more robots there are, the more likely robots will minimize time traveling in the MPFA. On the other hand, this makes the search time has smaller difference (see Fig.~\ref{fig:scalability_search_time}).  

It is obvious that the addition of more nests makes the travel time less for the MPFA. However, the information (pheromone waypoints) is distributed to multiple nests. In contrast to the CPFA, pheromone waypoints are globally available to all robots. So, there are tradeoffs among communication (and therefore search time) and travel time and congestion. In addition, the MPFA may get the benefit from all resources are not eventually be moved to one nest. However, we can consider a "high-speed" delivery (multiple targets can be moved in one round) in the future. This may not make the results too much different.
        
      
The search time for the MPFA decreases faster with increasing numbers of targets (see Fig.~\ref{fig:adaptation_search_time}). This shows that the MPFA has the advantage of reducing search time in an environment with large number of targets. We hypothesize that the evolved probability of laying pheromone increases and it is higher for the MPFA, or the rate of pheromone decay decreases and it is lower for the MPFA as the number of targets increases. The more targets there are, the more likely pheromone will be laid, or slower pheromone decay.  
 
These discoveries reveal that the MPFA improves foraging rates when the swarm size or the number of targets are scaled up. This is not only because of the simple intuitive reduction in travel time, but also because of the significant improvement in avoiding collisions. Overall, the MPFA has better performance as the swarm size or the number of targets increases.  

In the future work, we will discover the trends on the evolved seven parameters and confirm the above hypothesis for the random, partially clustered and clustered resource distributions. In addition, we will consider the cost of deploying multiply nests and evolve the optimized number of nests for different resource distributions.  






\section*{ACKNOWLEDGMENT}

We gratefully acknowledge members of the Moses Biological Computation Lab for their assistance with the multiple-place foraging swarm robotics project. Thanks to Antonio Griego for developing the CPFA algorithm in ARGoS. Thanks to Matthew Fricke for discussing the issues in the experiments.   



\end{document}